\documentclass[12pt]{article}
\begin{document}

\begin{center}
{\Large\bf{}An alternative non-negative gravitational energy
tensor to the Bel-Robinson tensor}
\end{center}

\begin{center}
Lau Loi So\\
Department of Physics, Tamkang University, Tamsui 251, Taiwan\\
(Dated on 30 Jan 2009,\quad{}s0242010@cc.ncu.edu.tw)
\end{center}

\begin{abstract}
The Bel-Robinson tensor $B_{\alpha\beta\mu\nu}$ gives a positive
definite gravitational energy in the small sphere limit
approximation. However, there is an alternative tensor
$V_{\alpha\beta\mu\nu}$ which was proposed recently that offers
the same positivity as $B_{\alpha\beta\mu\nu}$ does. These two
tensors are a basis for expressions which have the desirable
non-negative gravitational energy in the small sphere region
limit.
\end{abstract}

\section{Introduction}

The Bel-Robinson tensor $B_{\alpha\beta\mu\nu}$ has many nice
properties. It is completely symmetric, completely trace free and
completely divergence free.  It is usually regarded as being
related to gravitational energy.  In particular, the gravitational
energy-momentum density in the small sphere vacuum limit is
generally expected to be proportional to the Bel-Robinson tensor.
This expectation is related to the requirement of energy
positivity \cite{Szabados}.

However, we recently found another tensor \cite{SoarXiv},
$V_{\alpha\beta\mu\nu}$, which is also quadratic in the curvature,
and which enjoys the same positivity properties as
$B_{\alpha\beta\mu\nu}$. Moreover, we found that
$B_{\alpha\beta\mu\nu}$ and $V_{\alpha\beta\mu\nu}$ are a basis
for expressions which have the desirable non-negative
gravitational energy in the small sphere vacuum limit. In this
work, we examine some properties of $V_{\alpha\beta\mu\nu}$ and
some other quadratic in curvature tensors,
$S_{\alpha\beta\mu\nu}$, $K_{\alpha\beta\mu\nu}$ and
$W_{\alpha\beta\mu\nu}$, which have shown up in the expansion of
energy in the small sphere limit.

We found that $V_{\alpha\beta\mu\nu}$ fulfills the weak energy
condition in the small sphere limit.  We found another
$V'_{\alpha\beta\mu\nu}$, which does not satisfy the pseudotensor
conservation of energy-momentum restriction, but does satisfy the
weak energy condition.

The gravitational energy expression in the small region limit can
be investigated through the pseudotensors. In general, the
expansion of the pseudotensor expression can be represented by the
tensors $B_{\alpha\beta\mu\nu}$, $S_{\alpha\beta\mu\nu}$ and
$K_{\alpha\beta\mu\nu}$ if we consider the second order terms
\cite{SoarXiv,MTW}. Even though a pseudotensor is not a tensorial
object, this does not imply that it is useless. The second order
expansion expression provides guidance whether the gravitational
energy expression is positive or not \cite{SoarXiv}.

For the zeroth order term, the pseudotensor gives the mass density
as the equivalence principle demands. The non-vanishing second
order terms contribute the gravitational energy-momentum in a
small region limit; these terms are quadratic in the curvature
tensor.

\section{Quadratic curvature tensors}
There are three basic tensors that commonly occur in the
gravitational pseudotensor expression \cite{MTW,SoIJMPD}
\begin{eqnarray}
B_{\alpha\beta\mu\nu}
&:=&R_{\alpha\lambda\mu\sigma}R_{\beta}{}^{\lambda}{}_{\nu}{}^{\sigma}
+R_{\alpha\lambda\nu\sigma}R_{\beta}{}^{\lambda}{}_{\mu}{}^{\sigma}
-\frac{1}{8}g_{\alpha\beta}g_{\mu\nu}\mathbf{R}^{2},\\
S_{\alpha\beta\mu\nu}
&:=&R_{\alpha\mu\lambda\sigma}R_{\beta\nu}{}^{\lambda\sigma}
+R_{\alpha\nu\lambda\sigma}R_{\beta\mu}{}^{\lambda\sigma}
+\frac{1}{4}g_{\alpha\beta}g_{\mu\nu}\mathbf{R}^{2},\\
K_{\alpha\beta\mu\nu}
&:=&R_{\alpha\lambda\beta\sigma}R_{\mu}{}^{\lambda}{}_{\nu}{}^{\sigma}
+R_{\alpha\lambda\beta\sigma}R_{\nu}{}^{\lambda}{}_{\mu}{}^{\sigma}
-\frac{3}{8}g_{\alpha\beta}g_{\mu\nu}\mathbf{R}^{2},
\end{eqnarray}
where $\mathbf{R}^{2}=R_{\rho\tau\xi\kappa}R^{\rho\tau\xi\kappa}$.
Some properties of $S_{\alpha\beta\mu\nu}$ and
$K_{\alpha\beta\mu\nu}$ \cite{SoarXiv} are
\begin{eqnarray}
&&S_{\alpha\beta\mu\nu}\equiv{}S_{(\alpha\beta)(\mu\nu)}\equiv{}S_{\mu\nu\alpha\beta},\quad\quad{}
S_{\alpha\beta\nu}{}^{\nu}\equiv{}\frac{3}{2}g_{\alpha\beta}\mathbf{R}^{2}
,\quad\quad~{}S_{\alpha\mu\beta}{}^{\mu}\equiv{}0,\\
&&K_{\alpha\beta\mu\nu}\equiv{}K_{(\alpha\beta)(\mu\nu)}\equiv{}K_{\mu\nu\alpha\beta},\quad{}
K_{\alpha\beta\nu}{}^{\nu}\equiv{}-\frac{3}{2}g_{\alpha\beta}\mathbf{R}^{2},
\quad{}K_{\alpha\mu\beta}{}^{\mu}\equiv{}0.
\end{eqnarray}
Note that unlike $B_{\alpha\beta\mu\nu}$, both
$S_{\alpha\beta\mu\nu}$ and $K_{\alpha\beta\mu\nu}$ are neither
totally symmetric nor totally trace free.

For the quadratic curvature tensors, there are 4 independent basis
\cite{Deser} expressions, we may use
\begin{eqnarray}
\widetilde{B}_{\alpha\beta\mu\nu}
&:=&R_{\alpha\lambda\mu\sigma}R_{\beta}{}^{\lambda}{}_{\nu}{}^{\sigma}
+R_{\alpha\lambda\nu\sigma}R_{\beta}{}^{\lambda}{}_{\mu}{}^{\sigma}
=B_{\alpha\beta\mu\nu}+\frac{1}{8}g_{\alpha\beta}g_{\mu\nu}\mathbf{R}^{2},\label{20aJan2009}\\
\widetilde{S}_{\alpha\beta\mu\nu}
&:=&R_{\alpha\mu\lambda\sigma}R_{\beta\nu}{}^{\lambda\sigma}
+R_{\alpha\nu\lambda\sigma}R_{\beta\mu}{}^{\lambda\sigma}
=S_{\alpha\beta\mu\nu}-\frac{1}{4}g_{\alpha\beta}g_{\mu\nu}\mathbf{R}^{2},\\
\widetilde{K}_{\alpha\beta\mu\nu}
&:=&R_{\alpha\lambda\beta\sigma}R_{\mu}{}^{\lambda}{}_{\nu}{}^{\sigma}
+R_{\alpha\lambda\beta\sigma}R_{\nu}{}^{\lambda}{}_{\mu}{}^{\sigma}
=K_{\alpha\beta\mu\nu}+\frac{3}{8}g_{\alpha\beta}g_{\mu\nu}\mathbf{R}^{2},\\
\widetilde{T}_{\alpha\beta\mu\nu}
&:=&-\frac{1}{8}g_{\alpha\beta}g_{\mu\nu}\mathbf{R}^{2}.\label{20bJan2009}
\end{eqnarray}
These four tensors fulfill the symmetry
$\widetilde{M}_{\alpha\beta\mu\nu}=\widetilde{M}_{(\alpha\beta)(\mu\nu)}
=\widetilde{M}_{\mu\nu\alpha\beta}$. Although there exists some
other tensors different from $\widetilde{B}_{\alpha\beta\mu\nu}$,
$\widetilde{S}_{\alpha\beta\mu\nu}$,
$\widetilde{K}_{\alpha\beta\mu\nu}$ and
$\widetilde{T}_{\alpha\beta\mu\nu}$, they are just linear
combinations of these four.  For instance
\begin{equation}
\widetilde{T}_{\alpha\mu\beta\nu}+\widetilde{T}_{\alpha\nu\beta\mu}
\equiv\widetilde{B}_{\alpha\beta\mu\nu}
+\frac{1}{2}\widetilde{S}_{\alpha\beta\mu\nu}
-\widetilde{K}_{\alpha\beta\mu\nu}
+2\widetilde{T}_{\alpha\beta\mu\nu}.\label{7Nov2008}
\end{equation}
The above equality can be obtained by making use of the completely
symmetric property of the Bel-Robinson tensor. Using
(\ref{7Nov2008}), we can rewrite the Bel-Robinson tensor in a
different representation \cite{Deser}:
\begin{equation}
B_{\alpha\beta\mu\nu}\equiv-\frac{1}{2}S_{\alpha\beta\mu\nu}
+K_{\alpha\beta\mu\nu}+\frac{5}{8}g_{\alpha\beta}g_{\mu\nu}\mathbf{R}^{2}
-\frac{1}{8}(g_{\alpha\mu}g_{\beta\nu}+g_{\alpha\nu}g_{\beta\mu})\mathbf{R}^{2}.
\label{29Oct2008}
\end{equation}
This equation will be used in the next section.

\section{An alternative non-negative gravitational energy tensor}
Using the Riemann normal coordinate Taylor series expansion,
consider all the possible combinations of the small region
energy-momentum density in vacuum, the pseudotensor has the form
\cite{SoCQG}
\begin{equation}
2\kappa\,t_{\alpha}{}^{\beta}=2G_{\alpha}{}^{\beta
}+\left(\widetilde{a}_{1}\widetilde{B}_{\alpha}{}^{\beta}{}_{\mu\nu}
+\widetilde{a}_{2}\widetilde{S}_{\alpha}{}^{\beta}{}_{\mu\nu}
+\widetilde{a}_{3}\widetilde{K}_{\alpha}{}^{\beta}{}_{\mu\nu}
+\widetilde{a}_{4}\widetilde{T}_{\alpha}{}^{\beta}{}_{\mu\nu}
\right)x^{\mu}x^{\nu}+{\cal{}O}(\mbox{Ricci},x)+{\cal{}O}(x^{3}),
\label{25Sep2008}
\end{equation}
where $\kappa=8\pi{}G/c^{4}$ (with $c=1$ for simplicity) and
$\widetilde{a}_{1}$ to $\widetilde{a}_{4}$ are real numbers. From
now on, the second order term will be kept but the others are
dropped, because we are mainly interested in the gravitational
energy. The essential purpose of the present paper is to prove
that $B_{\alpha\beta\mu\nu}$ and $V_{\alpha\beta\mu\nu}$ are a
basis for positive gravitational energy in the small sphere limit.
There are two physical conditions which can constrain the
unlimited combinations between
$\widetilde{B}_{\alpha\beta\mu\nu}$,
$\widetilde{S}_{\alpha\beta\mu\nu}$,
$\widetilde{K}_{\alpha\beta\mu\nu}$ and
$\widetilde{T}_{\alpha\beta\mu\nu}$.  The first one is the
conservation of the energy-momentum density and the second is the
positive
gravitational energy in the small sphere limit.\\
$(\mathbf{i})$ First condition: energy-momentum conservation.
Consider (\ref{25Sep2008}) as follows
\begin{equation}
0=\partial_{\beta}\,t_{\alpha}{}^{\beta}
=\frac{1}{4}(\widetilde{a}_{1}-2\widetilde{a}_{2}+3\widetilde{a}_{3}
-\widetilde{a}_{4})g_{\alpha\beta}x^{\beta}\mathbf{R}^{2}.
\end{equation}
Therefore, the constraint for the conservation of the
energy-momentum density is
\begin{equation}
\widetilde{a}_{1}-2\widetilde{a}_{2}+3\widetilde{a}_{3}-\widetilde{a}_{4}=0.\label{9Jan2009}
\end{equation}
Although there are an infinite number of combinations which can
fulfill the above constraint, it has removed one degree of
freedom. As each single tensor of
$\widetilde{B}_{\alpha\beta\mu\nu}$,
$\widetilde{S}_{\alpha\beta\mu\nu}$,
$\widetilde{K}_{\alpha\beta\mu\nu}$ or
$\widetilde{T}_{\alpha\beta\mu\nu}$ cannot satisfy the
conservation requirement, but the sum with the others do.  One can
simplify the situation to eliminate
$\widetilde{T}_{\alpha\beta\mu\nu}$ which is being absorbed by
$\widetilde{B}_{\alpha\beta\mu\nu}$,
$\widetilde{S}_{\alpha\beta\mu\nu}$ and
$\widetilde{K}_{\alpha\beta\mu\nu}$. Consequently there are only 3
basis tensors left.  Rewrite (\ref{25Sep2008}) as
\begin{equation}
2\kappa\,t_{\alpha}{}^{\beta}=\left(a_{1}B_{\alpha}{}^{\beta}{}_{\mu\nu}
+a_{2}S_{\alpha}{}^{\beta}{}_{\mu\nu}
+a_{3}K_{\alpha}{}^{\beta}{}_{\mu\nu}
\right)x^{\mu}x^{\nu},\label{25aSep2008}
\end{equation}
where $a_{1}$ to $a_{3}$ are constants.  This is the reason why
these three tensors in (\ref{25aSep2008}) always appear to this
order when one investigates the gravitational energy using the
pseudotensor \cite{PRL}. Now, every tensor of
$B_{\alpha\beta\mu\nu}$, $S_{\alpha\beta\mu\nu}$ and
$K_{\alpha\beta\mu\nu}$ satisfies the condition of the
energy-momentum density conservation.\\
$(\mathbf{ii})$ Second condition: non-negative gravitational
energy in the small sphere limit. The purpose of the pseudotensor
is for determining the gravitational energy-momentum, the
associated energy-momentum can be calculated as
\begin{eqnarray}
2\kappa\,P_{\mu}&=&\int_{t=0}t^{\rho}{}_{\mu\xi\kappa}x^{\xi}x^{\kappa}d\Sigma_{\rho}
~=~t^{0}{}_{\mu{}lm}\int_{t=0}x^{l}x^{m}d^{3}x\nonumber\\
&=&t^{0}{}_{\mu{}lm}\frac{\delta^{lm}}{3}\int{}r^{2}d^{3}x
~=~t^{0}{}_{\mu{}l}{}^{l}\,\frac{4\pi{}r^{5}}{15},\label{5Jan2009}
\end{eqnarray}
where $l,m=1,2,3$. Using this calculation method, the
energy-momentum in the small sphere limit for (\ref{25aSep2008})
becomes
\begin{equation}
2\kappa\,P_{\mu}=(-E,\vec{P})=-\frac{r^{5}}{60}\left(
a_{1}B_{\mu{}0l}{}^{l}+a_{2}S_{\mu{}0l}{}^{l}+a_{3}K_{\mu{}0l}{}^{l}\right).\label{25dSep2008}
\end{equation}
The ``energy-momentum" values associated with
$B_{\alpha\beta\mu\nu}$, $S_{\alpha\beta\mu\nu}$ and
$K_{\alpha\beta\mu\nu}$  are proportional to
\begin{eqnarray}
B_{\mu{}0l}{}^{l}&=&(E_{ab}E^{ab}+H_{ab}H^{ab},2\epsilon_{c}{}^{ab}E_{ad}H^{d}{}_{b}),\\
S_{\mu{}0l}{}^{l}&=&-10(E_{ab}E^{ab}-H_{ab}H^{ab},0),\label{25bSep2008}\\
K_{\mu{}0l}{}^{l}&=&B_{\mu{}0l}{}^{l}-S_{\mu{}0l}{}^{l}.\label{25cSep2008}
\label{25cSep2008}
\end{eqnarray}
where the electric part $E_{ab}$ and magnetic part $H_{ab}$ are
defined in terms of the Weyl tensor as follows:
\begin{equation}
E_{ab}:=C_{a0b0},\quad{}H_{ab}:=*C_{a0b0}.
\end{equation}

Referring to (\ref{25aSep2008}), we are interested in the positive
gravitational energy within a small sphere limit, the Bel-Robinson
tensor already satisfies this condition. Precisely
\begin{equation}
B_{00l}{}^{l}=E_{ab}E^{ab}+H_{ab}H^{ab}\geq{}0.
\end{equation}
The rest of the job is to find the coefficients $a_{2}$ and
$a_{3}$.  Using (\ref{25cSep2008}), rewrite (\ref{25dSep2008}) as
\begin{equation}
2\kappa\,P_{\mu}=2\kappa\,(-E,\vec{P})=-\frac{r^{5}}{60}\left[
(a_{1}+a_{3})B_{\mu{}0l}{}^{l}+(a_{2}-a_{3})S_{\mu{}0l}{}^{l}\right].\label{27Sep2008}
\end{equation}
Equation (\ref{25bSep2008}) shows that $S_{\mu{}0l}{}^{l}$ cannot
ensure positivity, since we should allow for any magnitude of
$||E_{ab}||$ and $||H_{ab}||$. In other words, for
$S_{\alpha\beta\mu\nu}$ the sign of the ``energy" density is
uncertain. Therefore the only possibility for (\ref{27Sep2008}) to
guarantee positivity is when $a_{2}=a_{3}$. Recall the new tensor
$V_{\alpha\beta\mu\nu}$ \cite{SoarXiv} which is defined as
\begin{equation}
V_{\alpha\beta\mu\nu}:=S_{\alpha\beta\mu\nu}+K_{\alpha\beta\mu\nu}.
\end{equation}
Additionally, referring to (\ref{25cSep2008}),
\begin{equation}
V_{\mu{}0l}{}^{l}=S_{\mu{}0l}{}^{l}+K_{\mu{}0l}{}^{l}=B_{\mu{}0l}{}^{l}.
\end{equation}
Consequently (\ref{25dSep2008}) becomes
\begin{equation}
2\kappa\,P_{\mu}=-\frac{r^{5}}{60}\left(
a_{1}B_{\mu{}0l}{}^{l}+a_{2}V_{\mu{}0l}{}^{l}\right)
=-\frac{r^{5}}{60}(a_{1}+a_{2})B_{\mu{}0l}{}^{l}.
\end{equation}
Hence the proof is completed.  Indeed $B_{\alpha\beta\mu\nu}$ and
$V_{\alpha\beta\mu\nu}$ are a basis for expressions which have
non-negative gravitational ``energy" density in vacuum.

Note that $B_{\alpha\beta\mu\nu}$ and $V_{\alpha\beta\mu\nu}$ are
different tensors since they are defined by different fundamental
quadratic curvatures, explicitly
\begin{eqnarray}
B_{\alpha\beta\mu\nu}&=&\widetilde{B}_{\alpha\beta\mu\nu}+\widetilde{T}_{\alpha\mu\nu},\\
V_{\alpha\beta\mu\nu}&=&\widetilde{S}_{\alpha\beta\mu\nu}
+\widetilde{K}_{\alpha\beta\mu\nu}+\widetilde{T}_{\alpha\mu\nu}.\label{11Jan2009}
\end{eqnarray}
In particular $V_{\alpha\beta\mu\nu}$ is totally trace free but
not totally symmetric.  The following list shows some properties
\begin{eqnarray}
&&V_{\alpha\beta\mu\nu}\equiv{}V_{(\alpha\beta)(\mu\nu)}\equiv{}V_{\mu\nu\alpha\beta},\quad{}
V_{\alpha\beta\mu}{}^{\mu}\equiv{}0\equiv{}V_{\alpha\mu\beta}{}^{\mu},\\
&&V_{0000}\equiv{}V_{00l}{}^{l}\equiv{}V_{m}{}^{m}{}_{l}{}^{l}\equiv{}V_{ml}{}^{ml}
\equiv{}E_{ab}E^{ab}+H_{ab}H^{ab}\equiv{}B_{0000},\\
&&V_{\mu{}000}\equiv{}V_{\mu{}0l}{}^{l}\equiv{}V_{\mu{}l0}{}^{l}
\equiv{}(E_{ab}E^{ab}+H_{ab}H^{ab},2\epsilon_{c}{}^{ab}E_{ad}H^{d}{}_{b})\equiv{}B_{\mu{}0l}{}^{l}.
\end{eqnarray}

It is known that $B_{\alpha\beta\mu\nu}$ has the dominant energy
property \cite{Penrose}
\begin{equation}
B_{\alpha\beta\mu\nu}\,w_{1}^{\alpha}\,w_{2}^{\beta}\,w_{3}^{\mu}\,w_{4}^{\nu}\geq{}0,
\label{13Jan2009}
\end{equation}
where $w_{1}, w_{2}, w_{3}, w_{4}$ are any future-pointing causal
vectors. Using (\ref{29Oct2008}), rewrite $V_{\alpha\beta\mu\nu}$
as
\begin{equation}
V_{\alpha\beta\mu\nu}:=B_{\alpha\beta\mu\nu}+W_{\alpha\beta\mu\nu},
\end{equation}
where
\begin{equation}
W_{\alpha\beta\mu\nu}:=\frac{3}{2}S_{\alpha\beta\mu\nu}
-\frac{5}{8}g_{\alpha\beta}g_{\mu\nu}\mathbf{R}^{2}
+\frac{1}{8}(g_{\alpha\mu}g_{\beta\nu}+g_{\alpha\nu}g_{\beta\mu})\mathbf{R}^{2}.
\end{equation}
This tensor has some interesting properties
\begin{equation}
W_{\alpha\beta\mu\nu}u^{\alpha}t^{\beta}t^{\mu}t^{\nu}=0,\quad{}
W_{\alpha\beta\mu\nu}u^{\alpha}u^{\beta}u^{\mu}u^{\nu}=0,
\end{equation}
where $t$ is a timelike unit normal vector and $u$ can be timelike
or null.  $V_{\alpha\beta\mu\nu}$ contains more information than
$B_{\alpha\beta\mu\nu}$, however it seems that
$B_{\alpha\beta\mu\nu}$ is the important part of
$V_{\alpha\beta\mu\nu}$ and $W_{\alpha\beta\mu\nu}$ is a kind of
gauge freedom (i.e., it has no important physical effect).

A physical reasonable energy-momentum tensor has to fulfill the
energy condition.  The local energy density measured by the
observer with a 4-velocity should be non-negative.  The energy
condition must be true for all timelike unit normal vectors
\cite{Stephani}. In fact, we found $V_{\alpha\beta\mu\nu}$ has the
non-negative ``energy" property
\begin{equation}
V_{\alpha\beta\mu\nu}t^{\alpha}t^{\beta}t^{\mu}t^{\nu}
\equiv{}B_{\alpha\beta\mu\nu}t^{\alpha}t^{\beta}t^{\mu}t^{\nu}
=E_{ab}E^{ab}+H_{ab}H^{ab}\geq{}0.
\end{equation}
This is called the weak energy condition.  Because of the
continuity \cite{Stephani}, the above inequalities must still be
true if the timelike vector $t$ is replaced by a null vector $v$.
Indeed, we found
\begin{equation}
V_{\alpha\beta\mu\nu}v^{\alpha}v^{\beta}v^{\mu}v^{\nu}
\equiv{}B_{\alpha\beta\mu\nu}v^{\alpha}v^{\beta}v^{\mu}v^{\nu}\geq{}0.
\end{equation}
Therefore the statement is correct according to \cite{Stephani}
for $V_{\alpha\beta\mu\nu}$, which is based on the fact that
$B_{\alpha\beta\mu\nu}$ has the dominant energy property.

Following from (\ref{5Jan2009}), the energy-momentum density for
$V_{\alpha\beta\mu\nu}$ in the small sphere limit is
\begin{equation} 2\kappa\,P_{\mu}
=\frac{4\pi{}r^{5}}{15}\left(V^{0}{}_{\mu\alpha}{}^{\alpha}-V^{0}{}_{\mu0}{}^{0}\right)
=-\frac{4\pi{}r^{5}}{15}V_{0\mu{}00}.
\end{equation}
Or, more covariantly,
\begin{equation}
2\kappa\,P_{\mu}u^{\mu}
=-\frac{4\pi{}r^{5}}{15}\,V_{\mu\alpha\beta\gamma}u^{\mu}t^{\alpha}t^{\beta}t^{\gamma},
\end{equation}
where
\begin{equation}
V_{\alpha\beta\mu\nu}t^{\beta}t^{\mu}t^{\nu}\equiv{}
B_{\alpha\beta\mu\nu}t^{\beta}t^{\mu}t^{\nu}
=(E_{ab}E^{ab}+H_{ab}H^{ab},2\epsilon_{c}{}^{ab}E_{ad}H^{d}{}_{b}).
\end{equation}
The physical meaning (non-spacelike energy-momentum) is here
simpler and clearer than that of the dominant energy condition
(\ref{13Jan2009}). Obviously $V_{\alpha\beta\mu\nu}$ can play the
same role as $B_{\alpha\mu\nu}$, it ensures positivity in the
small sphere limit.  In other words, the ``energy-momentum"
density according to $B_{\alpha\beta\mu\nu}$ and
$V_{\alpha\beta\mu\nu}$ are on equal footing at the small sphere
region limit.

Moreover, from the technical point of view, if we are just
interested in the positive energy and relax the restriction on the
pseudotensor constraint, which means the conservation of the
energy-momentum, there are an infinite number of combinations that
have the weak energy condition, not including
$B_{\alpha\beta\mu\nu}$. We define
\begin{equation}
V'_{\alpha\beta\mu\nu}=\widetilde{K}_{\alpha\beta\mu\nu}+s\widetilde{S}_{\alpha\beta\mu\nu}
+t_{1}\widetilde{T}_{\alpha\beta\mu\nu}+t_{2}\widetilde{T}_{\alpha\mu\beta\nu}
+t_{3}\widetilde{T}_{\alpha\nu\beta\mu}.
\end{equation}
where $s, t_{1}, t_{2}, t_{3}$ are real numbers and
$t_{1}+t_{2}+t_{3}=1$.  Note that the energy-momentum contribution
for $\widetilde{S}_{\alpha\beta\mu\nu}$ can be ignored simply
because
\begin{equation}
\widetilde{S}_{\alpha\beta\mu\nu}u^{\alpha}t^{\beta}t^{\mu}t^{\nu}
=0=\widetilde{S}_{\alpha\beta\mu\nu}u^{\alpha}u^{\beta}u^{\mu}u^{\nu}.
\end{equation}
On the other hand,
\begin{equation}
\widetilde{T}_{\alpha\beta\mu\nu}u^{\alpha}u^{\beta}u^{\mu}u^{\nu}
\equiv\widetilde{T}_{\alpha\mu\beta\nu}u^{\alpha}u^{\beta}u^{\mu}u^{\nu}
\equiv\widetilde{T}_{\alpha\nu\beta\mu}u^{\alpha}u^{\beta}u^{\mu}u^{\nu}.
\end{equation}
Once again, $u$ can be timelike or null.  Similarly, from the weak
energy condition and using the continuity property, we found the
results of $V'_{\alpha\beta\mu\nu}$ as follows
\begin{eqnarray}
V'_{\alpha\beta\mu\nu}t^{\alpha}t^{\beta}t^{\mu}t^{\nu}
&\equiv&B_{\alpha\beta\mu\nu}t^{\alpha}t^{\beta}t^{\mu}t^{\nu}\geq{}0,\\
V'_{\alpha\beta\mu\nu}v^{\alpha}v^{\beta}v^{\mu}v^{\nu}
&\equiv&B_{\alpha\beta\mu\nu}v^{\alpha}v^{\beta}v^{\mu}v^{\nu}\geq{}0.
\end{eqnarray}
This illustrates that there exists an infinite number of other
tensors which have positivity if we exclude the conservation of
the energy-momentum density requirement according to the
pseudotensor restriction.

Furthermore, in order to obtain the dominant energy condition,
$B_{\alpha\beta\mu\nu}$ is the unique tensor that has the suitable
combination according to the 4 fundamental quadratic curvature
combinations, namely from (\ref{20aJan2009}) to
(\ref{20bJan2009}). The Bel-Robinson tensor has more nice
properties than the other quadratic curvature combinations
generally. However, concerning gravitational energy at the small
sphere limit, $V_{\alpha\beta\mu\nu}$ becomes the only alternative
choice to compare with $B_{\alpha\beta\mu\nu}$.

\section{Conclusion}
According to the four fundamental quadratic curvature tensors
\cite{Deser}, we construct all the possible combinations in the
pseudotensor expression.  We recovered that
$B_{\alpha\beta\mu\nu}$ gives a definite positive gravitational
energy in the small sphere limit approximation. However, we found
a unique alternative, the new tensor $V_{\alpha\beta\mu\nu}$,
which also contributes the same non-negative gravitational energy
density at the same region limit.  These two tensors can be
classified as a basis for expressions which have the desirable
non-negative gravitational energy in the small sphere region
limit. Moreover, we found that the tensor $W_{\alpha\beta\mu\nu}$,
associated with $V_{\alpha\beta\mu\nu}$, behaves as a kind of
gauge freedom.

Relaxing the restriction of the energy-momentum conservation
requirement for the pseudotensor, $V'_{\alpha\beta\mu\nu}$
demonstrates that there are an infinite number of ways to obtain
positivity, namely the weak energy condition. For the conserved
expressions $B_{\alpha\beta\mu\nu}$ satisfies the dominant energy
condition while $V_{\alpha\beta\mu\nu}$ satisfies the weak energy
condition.

\section*{Acknowledgment}
This work was supported by NSC 97-2811-M-032-007.



\end{document}